\documentstyle[11pt, newpasp, twoside, epsf]{article}
\begin{document}
\title{Some Common Problems and Challenges that Emission-Line
Stars Present}
\author{Michael Feast}
\affil{Department of Astronomy, University of Cape Town,
Rondebosch, 7701, South Africa. 
email: mwf@artemisia.ast.uct.ac.za}
\begin{abstract}
Four examples are given of problems that arise in the study of
emission-line stars, each of which deserves further quantitative
study. (1) Archival slit spectra of $\eta$ Car show that epochs of
low excitation have occurred regularly in the 2020 day period
of Damineli et al. since at least 1948. They last about 10 percent
of the period. Earlier slit spectra (1899 -1919) suggest that the
the object was always in a low excitation state at that time. This
result may be connected with the 
gradual brightening of $\eta$ Car through
most of the 20th century. 
(2) FeII and [FeII] emission in Mira variables is probably
chromospheric. It would be important to understand how this is
excited and its relation to mass loss and dust formation in Miras.
(3) Declines of RCB stars are due to eclipses by dust clouds. At such
times a rich emission line spectrum from the outer atmosphere of the
star is seen. The quantitative study of the changes in this
spectrum as more and more of the star is eclipsed should yield
important information on the extended atmosphere of these objects.
(4) In deep minima when the entire star is eclipsed, RCB stars show
a broad-line emission spectrum. The excitation of this spectrum, particularly
HeI 3888A, constitutes a puzzle. Two possible excitation
mechanisms are discussed.

\end{abstract}

\section{Introduction}
The study of emission-line stars presents many opportunities
to probe the physical nature of stellar atmospheres and 
stellar environments and also to investigate little
understood modes of stellar evolution. This has been the
theme of much of this meeting. In an attempt to show that
we have as yet only begun to unlock this rich source of
information, this paper begins with a discussion of results
obtained using archival spectra of $\eta$ Carinae and then
examines two classes of emission-line stars which have not
so far been mentioned at this meeting but which would
repay further quantitative study.

\section{Archival Spectra of $\eta$ Carinae}
Gaviola (1953) found that there was a marked difference between
spectra of $\eta$ Car taken in 1948 and those taken in 1947 and 1949.
HeI, [NeIII] and various other emission lines of above average
excitation were missing in 1948. A similar low excitation event
was found to occur in 1965 (Thackeray 1967, Rogers \& Searle 1967)
and another in 1981 (Whitelock et al. 1983, Zanella et al. 1984).
In the latter case the low excitation phase occurred
when the infrared ($JHKL$) flux was near a maximum in its $\sim$ 5 year
cycle. Damineli (1996) showed that these low excitation events
all occurred at the same phase in his proposed 5.52 year period.
As one part of a recent study of $\eta$ Car (Feast, Whitelock 
\& Marang 2000)
we have examined spectra in the archives of the South African
Astronomical Observatory (SAAO) for further evidence of variability.
The bulk of these spectra were taken by A.D. Thackeray between
1951 and 1978. He made a large number of observations of the
central object (as seen from the ground) and of portions of the
homunculus as well as other surrounding nebulosity. 
Although there may well be relatively subtle differences between
different spectra of the central object, including secular variations,
it proved quite easy to classify spectra either as in a high or a low
excitation state with only one, rather doubtful, intermediate case.
Thackeray's plates reveal three previously unrecorded epochs of
low excitation. Taken together with other evidence in the literature,
these results show (see Fig. 1) that epochs of low excitation have occurred
regularly in the 2020 day period proposed by Damineli et al. (2000),
at least since 1948. The low excitation lasts about 10 percent
of the period though there is probably some slight variation in
the phase of its onset and end. The epochs of low excitation have
the same phase and duration as the eclipse shown by the infrared
light curves (Whitelock 2001). It is known (Davidson et al. 1997)
that the (sharp) emission lines used to judge the high/low
excitation phase come from plasma blobs about 0.2 arcsec from the
central object itself. The association of the low excitation
phases with the eclipse phases then presumably indicates that
the relevant emission lines are radiatively excited and, on
a binary model, that the blobs lie in or near the orbital
plane of the binary. This is consistent with the blobs lying
near the equatorial plane of the bipolar lobes as found
by Davidson et al. (1997).\\

\begin{figure}
\epsfxsize=13.2cm
\epsffile{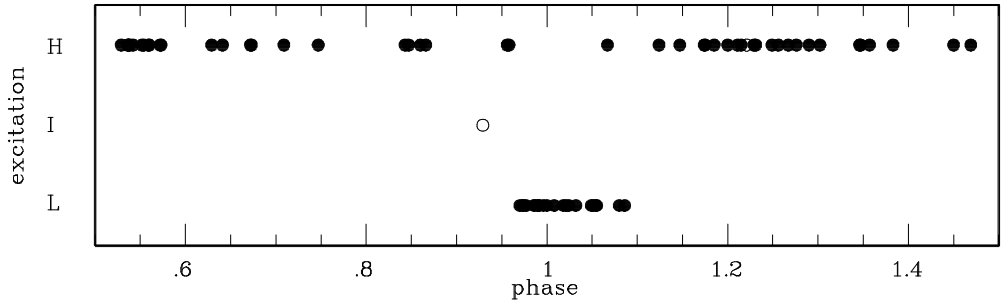}
\caption{The phases of high (H), low (L), and intermediate (I),
excitation in the 2020 day period since 1948. The open
circles represent the two less certain assignments.}
\end{figure}
Whilst there are earlier visual and objective prism observations of
$\eta$ Car, the first slit spectrum seems to be that take by Gill
at the Cape in 1899. This spectrum could not be located but
reproductions of it have been published
(Gill 1901). The Cape archives contain
slit spectra from 1919 
(Lunt 1919) and glass copies of the Lick Southern Station
plates of 1912-14
(Moore \& Sanford 1913 and unpublished). 
All the early spectra are in the low excitation
phase and suggest that the object was always in a low state at that time.
This result seems consistent with the evidence that the general level
of excitation is increasing with time (Whitelock et al. 1983, 
Damineli  et al. 1999). Presumably this is connected with the
secular increase in brightness of the system. Although this 
brightening has in
the past been ascribed by some workers to clearing of dust absorption, 
this seems unlikely to be the basic cause
(Whitelock et al. 1983, Davidson et al. 1999). 
One possibility is that the increase
in brightness and level of excitation is due to an increase in output
from the ``central engine''. On a binary model this could be due to
increased wind/disc or wind-wind interaction.\\

\section{Emission Lines in Mira Variables}
One can classify the optical emission lines seen in Mira variables into
four groups. 
(There are of course, in addition, maser and non-maser molecular
emission lines in the radio region.) 
First, there are the Balmer lines 
and a few others which must, at
least at some phases, originate within the stellar atmosphere
itself because
they are at times mutilated by overlying absorption. 
The evidence for this is rather conclusive (e.g. Merrill 1945, Joy 1947)
despite the claim that the resulting unusual Balmer decrement can
be alternatively explained by NLTE radiative transfer effects
(Luttermoser \& Bowen 1992).
These lines
are general believed to be excited by shock waves generated by
stellar pulsation. Secondly, there are what appear at first sight
to be a rather random selection of neutral metallic lines
(lines of FeI (e.g. 4202A), TiI and InI). These were explained
by Thackeray (1937) as due to fluorescent excitation (due
to chance coincidences of energy level differences) by lines
(principally MgII h and k)
which are presumably associated with the first group. Thirdly there are
molecular lines, including the fascinating case of AlH emission
by inverse pre-dissociation (Herbig 1956, Herbig \& Zappala 1968). 
Finally there are 
emission lines of FeII and [FeII] which can be found in some,
and probably all, large amplitude Miras near minimum light.
Along with these lines are semi-forbidden MgI] (4571A) and some other
forbidden lines ([OI], [SII]). The origin of the FeII and [FeII]
lines is particularly interesting. There seems to be no indication
that these lines are affected by overlying atmospheric absorption. 
This needs to be properly checked but it suggests that the lines
are excited in an outer region of the star. This is indeed what Herbig 
(1969) deduced from a relatively simple semi-quantitative analysis
of the line strengths. He found that the lines were excited in a
chromosphere at T $\sim$ 4000K, i.e. significantly hotter than the
photosphere. The excitation of chromospheres is a problem of
continuing interest. How do Miras do it? The most likely mechanism
is probably that pulsational shocks propagate into the outer
atmosphere and excite a chromosphere. Wood \& Karovska (2000)
have recently shown that the intensity of ultraviolet FeII emission
lines in Miras varies with phase, consistent with this hypothesis.
The case of possible chromospheric emission in Miras is of
particular interest because these stars are undergoing extensive
mass loss and dust formation. It is not clear how these various
processes are related although it is almost certainly misleading
to think of the outer region of a Mira as consisting of well
ordered layers. More likely is a low density region containing
denser clouds.
Chromospheric emission in Miras is particularly interesting
since in
late type giants and supergiants, in general, the strength of
chromospheric emission is reduced by the presence of dust
(see e.g. Stencel, Carpenter \& Hagen 1986).\\ 
 
\section{Emission Lines in R Coronae Borealis Variables (I)}
The emission-line spectra seen from time to time in RCB variables
raise a number of interesting problems which deserve more
extensive quantitative study. These stars are highly evolved
carbon-rich, hydrogen-poor objects
(see e.g. Clayton 1996, Feast 1996, and earlier reviews). Their
 evolutionary history is still uncertain. They are high luminosity
objects, typically with spectra similar to F or G-type supergiants.
But they are almost certainly low mass objects. At random intervals
RCB stars decrease rapidly in brightness by many magnitudes and then
slowly brighten again. This behaviour is attributed to the star
ejecting dense clouds of particles in random directions, one of
which occasionally falls in the line of sight and causes an eclipse.
It has been suggested (Asplund \& Gustafsson 
1996, Asplund 1998) that this behaviour
is a result of the stars being at an opacity-modified Eddington limit.
It may be recalled that the behaviour of the LBV variables, discussed
earlier in this meeting, has also been attributed  
(e.g. Appenzeller 1989) to their
being at an opacity-modified Eddington limit.
One way in which these objects differ is that whilst RCB stars puff
off clouds of limited size, the LBVs are thought to eject complete
shells. Asplund \& Gustafsson suggest that this difference is due
to the fact that the LBVs are evolving to the red. Thus any expansion
and cooling makes them more super-Eddington. The RCBs on the other
hand are probably evolving to the blue thus expansion and cooling will
tend to make them sub-Eddington. Asplund \& Gustafsson speculate
that this results in a local, rather than a general, instability
and gives rise to the ejection of limited clouds rather than 
complete shells.\\ 

Since much of this meeting has been devoted to
$\eta$ Car it is worth noting that Bidelman (1993) thought that
the 1893 (objective prism) absorption spectrum of $\eta$ Car
was similar to
that of R CrB at maximum light in showing strong CI absorption. 
If this is actually the case it would add to the puzzle provided by 
Lamers et al. (1998) 
who found that the carbon abundance in the stellar wind from 
$\eta$ Car is greater than in the surrounding nebulosity. Perhaps
this requires a two-wind model of some kind for $\eta$ Car. 
\\

Because the RCB eclipses are due to dust clouds they cannot
be expected to repeat exactly, being dependent on the size,
opacity etc. of the cloud. Nevertheless on some declines at least
a very rich emission-line spectrum is found (e.g. Alexander et al. 1972).
This is dominated by FeII and is, with some minor differences, a
reversal of the absorption spectrum at maximum light. Evidently
the main body of the star is eclipsed and we see the equivalent
of the solar-eclipse ``flash" spectrum. This spectrum changes, 
presumably as the eclipsing cloud expands and covers more and more of
the outer atmosphere of the star. From being initially dominated
by FeII emission, it changes to being dominated by TiII and ScII. It
has been qualitatively suggested that these changes are
due to a decrease in temperature, density and self-absorption
with height in the outer atmosphere but a full quantitative discussion
is required and might throw considerable light on this outer region which
probably has a diameter roughly four times that of the photosphere.
Note that these spectra are often referred to as ``chromospheric".
However there is no evidence 
that the emission lines come from a region above a temperature minimum.

\section{Emission Lines in R Coronae Borealis Stars II}
There is another, entirely different, problem concerning the
emission-line spectra of RCB stars which also raises a rather
interesting physical problem.\\

In deep minimum the sharp-line spectrum referred to in the last section
has completely disappeared. Presumably the star, including its outer
atmosphere, is totally eclipsed by a dust cloud. The spectrum
then consists of a small number of broad emission lines. The main lines
in the optical region are HeI 3888A, CaII H and K, and the NaI D lines.
Their width is $\sim$ 400 $\rm km\:s^{-1}$ (see, e.g. Alexander et al. 1972,
Rao et al. 1999). These lines are characteristic of all RCB stars that
have been suitably examined in deep minimum. They 
clearly originate far from the central star since they are
unobscured. But where exactly do they originate and how are the excited?
\\

One model
has been proposed by Rao et al. (1999). They suppose that R CrB
itself, and by implication all RCB stars, are binaries with 
white dwarf companions. The white dwarf, they suggest, is surrounded by
an accretion disc which is the source of the broad emission lines.
The width of the lines is due to the Keplerian rotation of the disc.
The HeI emission is supposed to come from the inner parts of the disc.
It is not then clear why the CaII and NaI lines have similar
width to the HeI since they presumably would arise in the cooler,
outer parts of the disc (or from elsewhere). This, together with
the fact that the model requires that all RCB stars are viewed
close to the plane of the supposed binary and the lack of any
direct evidence for a white dwarf in any of the variables, makes
the hypothesis rather unattractive.\\

There are in fact purely phenomenological considerations which
suggest a quite different model
(Feast 1996). When the dust cloud in front
of the star is thin enough to see through,
so that the star itself is dimly seen, one also detects absorption
lines at $\sim \rm -200\:km\:s^{-1}$. These lines must come
from gas entrained with the dust which is being driven from
the star at this speed by radiation pressure. These lines may
be seen at the start of a decline, as the dust thickens, or
on the rise back to maximum light as the dust cloud expands 
and thins (see, e.g. Alexander et al. 1972). Evidently we should
consider the star as being surrounded by a consortium of clouds, ejected
on a quasi-continuous basis in all directions at a typical
velocity of $\sim \rm 200\:km\:s^{-1}$.
The displaced lines found are not only CaII H and K, and the NaI
D lines (which are probably easily understood), but also HeI 10830A
(see, e.g. Querci \& Querci 1978, Feast 1986). This indicates
that these dust and gas clouds contain HeI in the highly
metastable $\rm ^{3}S$ state which is the lower level of both
the 10830A and the 3888A lines. This being the case it is clear
that when all but the consortium of clouds is obscured we would expect to
see, as is actually found, resonance emission lines of CaII and NaI
and also HeI lines 
resonantly excited from the
$\rm ^{3}S$ state.
The width of these lines would be $\rm \sim 2 \times 200\:km\:s^{-1}$
as is indeed found. The case that the broad emission lines
(including HeI 3888A) arise in the expanding dust/gas clouds
thus appears to be rather strong. The problem then is how
the $\rm ^{3}S$ HeI level, which is nearly 20 eV above the
ground state, is populated. This is a question which has not been
fully answered.
It may be that some
form of atomic collisions (Feast 1996) or shocks 
(as suggested by J. Linsky in the discussion at this meeting)
in the fast moving
dust/gas clouds could provide the necessary excitation. It should be
remembered that not only is helium the most abundant element
in RCB star atmospheres, but also that the metastability of
$\rm ^{3}S$ HeI ensures that 3888A (and 10830A) can be resonantly
excited many times from an atom in this state before it
returns to the ground state. Thus the rate at which HeI atoms
need to be raised to the $\rm ^{3}S$ state is likely to be quite
low. Progress is likely  to come from more quantitative observational
work, from theoretical studies and, one might guess, from laboratory
experiments. This may indeed be one of those areas where a solution
to a problem lies in the collaborative efforts of astronomers
and laboratory physicists.
 
\acknowledgments
I am grateful to Patricia Whitelock for data in advance of publication
and for many helpful discussions.

\end{document}